\documentclass[useAMS,usenatbib]{rspublic}
\usepackage{myaasmacros}
\input{epsf}

\def\ltsima{$\; \buildrel < \over \sim \;$}
\def\simlt{\lower.5ex\hbox{\ltsima}}   
\def\gtsima{$\; \buildrel > \over \sim \;$}
\def\simgt{\lower.5ex\hbox{\gtsima}}
\newcommand\bcite[1]{\citeauthor{#1} \citeyear{#1}}

\title[The baryonic mass function of galaxies]
{The baryonic mass function of galaxies}

\author[Read \& Trentham]{J. I. Read $^1$\thanks{Email: jir22@ast.cam.ac.uk} 
\& Neil Trentham $^1$ 
\\ $^1$Institute of Astronomy, Cambridge University, Madingley Road, 
Cambridge, CB3 OHA, England}

\begin{document}

\maketitle

\begin{abstract}{mass function, stellar, HI, gas, baryonic}

In the Big Bang about 5\% of the mass that was created was in the form
of normal baryonic matter (neutrons and protons). Of this about 10\%
ended up in galaxies in the form of stars or of gas (that can be in
molecules, can be atomic, or can be ionised).

In this work, we measure the baryonic mass function of galaxies, which 
describes how the baryonic mass is distributed {\it within} 
galaxies of different types (e.g.~spiral or elliptical) and of
different sizes. This can provide useful constraints on our
current cosmology, convolved with our understanding of how
galaxies form. This work relies on various large astronomical surveys,
for example the optical Sloan Digital Sky Survey (to observe stars) 
and the HIPASS radio survey (to observe atomic gas). We then perform
an integral over our mass function to determine the cosmological
density in baryons in galaxies: $\Omega_{\rm b, gal} = 0.0035$. 
Most of these baryons are in stars: $\Omega_{*} = 0.0028$.
Only about 20\% are in gas. The error on the quantities, as determined
from the range obtained between different methods, is
$\sim 10$\%; systematic errors may be much larger.

Most ($\sim 90$\%) of the baryons in the Universe are not in galaxies. They
probably exist in a warm/hot intergalactic medium. Searching for
direct observational evidence and deeper theoretical understanding for
this will form one of the major challenges for astronomy in the next
decade.
 
\end{abstract}

\section{Introduction}\label{sec:introduction}

The hot Big Bang theory is the current paradigm for the formation and evolution of the Universe.
Parts of the theory that describe the Universe in the first small fraction
of a second, like superstring theory,
cannot rigorously be tested by experiment and so are not well-constrained at present.
But the parts that describe the subsequent evolution of the Universe have been very successful.
Examples of the successes are the predictions of the cosmic microwave background (CMB), the abundances
of light elements and the expansion of the Universe
\citep{1999coph.book.....P}.

One feature of the model is the partitioning of mass-energy into entities 
that obey different equations of state.
In Table 1, we list current estimates for the densities of these forms 
of mass-energy (\citeauthor{2004ApJ...616..643F}~2004, 
\citeauthor{2004MNRAS.353..457A}~2004).
These numbers come from matching the theory to observations of the CMB
(e.g. \citeauthor{2003ApJS..148..175S}~2003), 
the distance-redshift relation obtained from 
Type Ia supernova (e.g.~\citeauthor{2004ApJ...607..665R}~2004),
and observations of rich galaxy clusters a different redshifts (Allen et al.~2004). 
The two main components are a dark energy component with density $\Omega_\Lambda$, which could be a
cosmological constant\footnote{The energy density of
a given component, $\Omega_i=\rho_i/\rho_c$, is defined as the ratio of
its density to the critical density which would make the universe
spatially flat: $\rho_c=3H_0^2/8\pi G$, where $G$ is the 
gravitational constant and $H_0$ is the Hubble constant
at the present epoch, which we assume to be
$H_0 =
70 $ km s$^{-1}$Mpc$^{-1}$.
The astronomical unit of a parsec (pc) is 3.09 $\times 10^{16}$ m.}, and a matter component
with density $\Omega_M$. 
The matter density can further be divided into a dark-matter component and a baryonic component.
Inferences about the density of the baryonic component come from observations of the CMB angular power 
spectrum \citep{2003ApJS..148..175S} and the abundances of light
elements D, $^{3}$He, $^{4}$He and $^{7}$Li
(\citeauthor{2004ApJ...600..544C}~2004) when compared with
big bang nucleosynthesis theory (BBN). The existence of the dark matter
component is strengthened by independent evidence from the dynamics
and clustering of galaxies and clusters of galaxies 
(e.g. \citeauthor{2001ApJ...552L..23D}~2001,
\citeauthor{2001ApJ...563L.115K}~2001 and
\citeauthor{2001MNRAS.323..285B}~2001); and from 
gravitational lensing measurements (\bcite{1999ARA&A..37..127M} and
\bcite{2004ApJ...604...88S}). The baryonic component is the
subject of the present paper.

\begin{table}
\caption{Composition of the Universe}
{\vskip 0.5mm}
{$$\vbox{
\halign {\hfil #\hfil && \quad \hfil #\hfil \cr
\noalign{\hrule \medskip}
 &  & & & & &\cr
\noalign{\smallskip \hrule \smallskip}
\cr
Total density & & Dark energy density& & & \cr
$\Omega=1$  & &$\Omega_\Lambda=0.73$ & & &\cr
              & & Matter density& & Dark matter density & \cr
              & & $\Omega_M=0.27$ & & $\Omega_{\rm dm}=0.22$ & & &\cr
              & &               & & Baryonic density & \cr
              & &                      & & $\Omega_{\rm b}=0.05$ & & &\cr
\noalign{\smallskip}\cr}}$$}
\vspace{-10mm}
\end{table}

Understanding the present distribution of material in the Universe
is a formidable challenge. With the advent
of fast computers and efficient algorithms for calculating the force
between many particles, we can now make solid predictions
for the current distribution of dark matter
(\citeauthor{1998ARA&A..36..599B}~1988). However, understanding the distribution
of stars and gas is more difficult. This involves a detailed
understanding of the physics underlying galaxy formation: gas
hydrodynamics, star formation, and feedback from exploding stars and
forming black holes. Despite these difficulties, much
progress has been made and simulations are now on the verge of being
able to make predictions about where and in what form we should expect
to find baryons\footnote{Baryons are any hadron which comprises three
  quarks; typically this refers to protons and neutrons.} in the
universe today (see e.g. \bcite{2004bdmh.conf...37M} and
\bcite{2005Natur.435..629S}). 
The new hydrodynamic simulations of galaxy formation
(e.g. \bcite{2001ApJ...552..473D} and
~\citeauthor{2005ApJ...618...23N}~2005)
include an intergalactic medium, which makes it straightforward to include
these effects.

Technological advances in observational astronomy have also been rapid, 
mainly as regards our ability to compile and process large datasets. The results from 
many surveys have recently been published. These include optical redshift 
surveys, like the Sloan Digital Sky Survey (SDSS; \citeauthor{2004AJ....128..502A}~2004)
and the 2DF Galaxy Redshift Survey \citep{2001MNRAS.328.1039C}, near-infrared 
photometric surveys like 2MASS (\citeauthor{2004PABei..22..275G}~2004), and HI atomic gas surveys 
like HIPASS (\citeauthor{2004AJ....128...16K}~2004).

The time, therefore, is ripe to use the
survey results in conjunction with each other 
to calculate precisely the distribution of
baryons in the Universe. This will then provide a valuable resource
against which cosmological simulations which include the complex
physics of star formation and galaxy formation may be tested. It also
presents a vital census of that part of the universe which we can see.
While models which invoke dark matter or dark energy may come and go,
the light we observe from galaxies and the distribution of the baryons
within these systems remains a firm result.
 
A first attempt at such a calculation was made by
\citet{1999MNRAS.309..923S} for disc galaxies 
while \citet{2003ApJ...585L.117B} recently presented the
baryonic mass function of all galaxies, calculated using a combination
of 2MASS and SDSS data. Other authors have
computed the stellar mass function of galaxies
(\citeauthor{2001MNRAS.326..255C}~2001 and
\citeauthor{2004MNRAS.355..764P}~2004) or the gas mass function 
(\citeauthor{2003ApJ...582..659K}~2003, Koribalski et al.~2004, 
\citeauthor{2004AAS...20512203S}~2004).

In this paper we present the baryonic mass
function for nearby galaxies separated by Hubble
Type\footnote{The Hubble `tuning fork' galaxy
  classification scheme separates galaxies by morphology into `early types'
  (elliptical and lenticular - E, S0) and `late types' (spirals - Sa,
  Sb, Sc, Sd, Sm -  and irregulars - Irr). Dwarf galaxies are denoted
  dE (dwarf elliptical) or dIrr (dwarf irregular). In this paper we
  group the dIrr and Irr galaxies together. Note that while such
  classifications are historical, these galaxies are genuinely distinct
  in their chemical and dynamical properties. E, dE and S0
  galaxies are largely supported by random stellar motions and are of
  typically higher metallicity, while
  Sa/b/c/d, Irr and dIrr are largely rotationally supported and of
  lower metallicity. The dynamical
  differences, plus the differences in chemistry make determinations of
  the stellar and gas mass-to-light ratios for these systems necessarily
  different. Thus splitting the galaxy populations into different types
  in this way is useful for a study such as this one.}.
We convert the the field galaxy luminosity function of
\citet{2005MNRAS.tmp...52T}
(which depends on SDSS measurements at the bright end)
to a mass function using dynamical mass estimates where possible
(e.g.~\citeauthor{2000AAS..144...53K}~2000) and stellar population synthesis
models otherwise (e.g.~\citeauthor{2003MNRAS.344.1000B}~2003). We use data from
the HIPASS survey (Koribalski et al.~2004)
to determine the HI (atomic hydrogen) gas masses of galaxies 
and data from the FCRAO Extragalactic CO survey of Young et al.~(1995) to 
determine the H$_2$ (molecular hydrogen) gas masses of galaxies
of a given luminosity and Hubble Type.
We further use recent X-ray surveys of disc and elliptical
galaxies to constrain the contribution of ionised hydrogen
\citep{2003ARAA..41..191M} and use microlensing data to
constrain the contribution of stellar remnants
(\citeauthor{2000ApJ...541..734A}~2000, \citeauthor{2001A&A...373..126D}~2001 and
\citeauthor{2003AA...404..145A}~2003). 
Combining all these measurements gives the baryonic mass function.

This paper is organised as follows. In section \ref{sec:luminosity} we
present the luminosity function of field galaxies separated by Hubble
Type. In section \ref{sec:stars} we describe our method for calculating
mass-to-light ratios of stellar populations in different galaxies and compute
the stellar mass function of galaxies. 
In section \ref{sec:gas} we compute the gas mass
function for field galaxies. In
section \ref{sec:baryons} we present the full baryonic mass function
of galaxies and compare this with results from the literature. 
We then discuss baryonic mass components which
have been left out of the analysis: halo stars, stellar remnants, 
and hot ionised gas and dust. In
section \ref{sec:omega} we integrate the baryonic mass function
to obtain the relative contribution of each baryonic mass component to
$\Omega_{\rm b, gal}$ and compare our results to values of 
$\Omega_{\rm b}$ derived from the CMB and light element abundances.
Finally, in section
\ref{sec:conclusions} we present our conclusions.

\section{The luminosity function of field
  galaxies}\label{sec:luminosity}

The luminosity function $\phi(M_R)$, is defined such that
$\phi(M_R)\,{\rm d}M_R\,{\rm dV}$ is the number of galaxies in the
absolute $R$-band magnitude range [$M_R$, $M_R + {\rm
  d}M_R$]\footnote{The absolute magnitude is a logarithmic measure of
  the luminosity in the $R$-band. Historically, the luminosity of
  galaxies is presented as $M_X=-2.5\log_{10} \int_{X} L_{\lambda} T_X
  ({\lambda}) \, {\rm d}{\lambda} +$constant where $X$ denotes some
  filter (e.g.~$R$-band), $T_X ({\lambda})$ is the transmission
  function of that filter, and $L_{\lambda}$ is the spectral energy
  distribution of the galaxy. The nomenclature for filters is
  complicated because many groups define their own system. Often these
  labels overlap so that the `r-band' has several different
  definitions in the literature. For a full review see
  \citet{1995PASP..107..945F}. In this paper we will refer only to the
  Cousins $R$-band (5804-7372 \AA) and the Johnson $B$-band (3944-4952
  \AA) (\citeauthor{1995PASP..107..945F}~1995). In the days of
  photographic astronomy, the names were often taken from the types of
  plates supplied by photographic companies. For example, the $J$ in
  the well-known $b_J$ filter comes from the particular type of
  photographic emulsion obtained from Kodak. We even learnt one story
  where the notation had to be changed when the photographic company
  changed the plate name because the yak whose stomach lining they
  used to manufacture the glue used in the emulsion became
  endangered!} within co-moving volume element dV. The field galaxy
luminosity function is plotted by Hubble Type in 
Figure 1. As in \citet{2005MNRAS.tmp...52T}, the bright end of the luminosity
function was computed using data from the SDSS galaxy survey, while the
faint end was taken from observations of nearby galaxy
groups (e.g.~\citeauthor{2002MNRAS.335..712T}~2002). We show
parameter values on the plot for a Schechter function fit to the
luminosity function\footnote{The Schechter function in mass ($M$) or
  luminosity ($L$) is given by:
  $\phi(M)=\phi_*\exp\left(-\frac{M}{M_*}\right)\left(\frac{M}{M_*}\right)^\alpha$. In magnitude units this gives: $\phi(M_R) = 0.92\phi_*\left(10^{[-0.4(M_R-M_{R_*})]}\right)^{\alpha+1}\times\exp\left(-10^{[-0.4(M_R-M_{R_*})]}\right)$ \citep{1976ApJ...203..297S}.}. While the
error bars are small enough that the Schechter function does not
provide a good formal fit, it is a simple analytic form which captures
the main features of the mass and luminosity functions which are
presented in this paper.

\begin{figure} 
\begin{center}
\epsfxsize=12truecm\epsfysize=12truecm\epsfbox{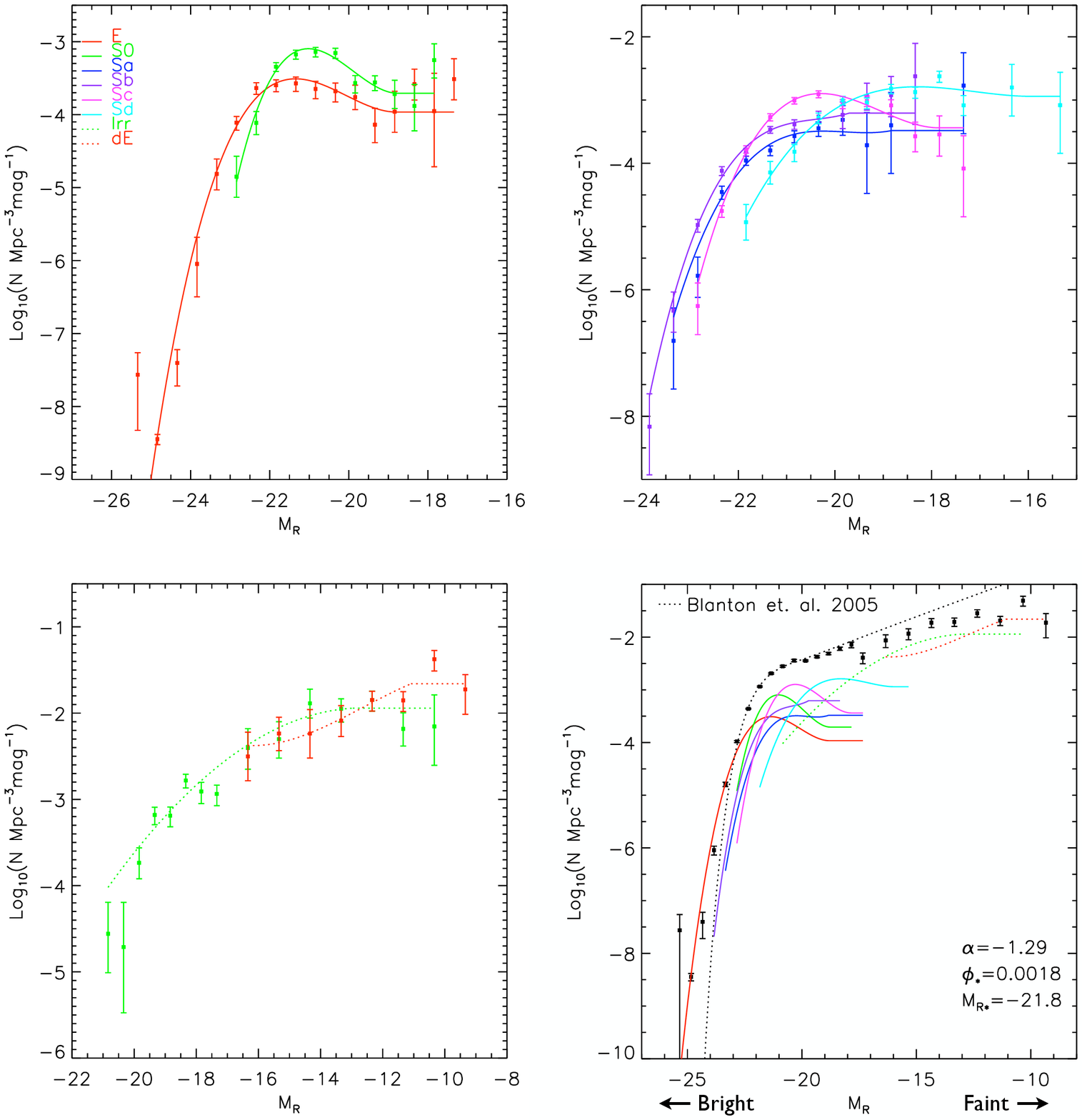}
\end{center}    
  \caption[]
  {The field galaxy luminosity function split by Hubble Type. The top
    left plot is for Elliptical and S0 galaxies, the top right is for
    spirals (Sa-Sd), the bottom left is for the dwarf galaxies
    (irregular and elliptical) and the bottom right is the combined
    luminosity function. In each plot a spline-fit to the luminosity
    function is shown to guide the eye and these spline fits are also
    overlaid on the combined luminosity function in the bottom right
    plot. Galaxies which are very luminous in the R-band lie to {\it
    left} of these plots, while those which are very faint lie to the
    right. Overlaid on the bottom right panel are parameters for a
    Schechter fit to the total luminosity function.}
\label{fig:lumfunction}
\end{figure}

The Hubble Type is a subjective assessment that depends on many
parameters that can be measured for nearby bright galaxies. It cannot
be determined for the  
faint galaxies in SDSS images, so we cannot determine type-specific luminosity 
functions from SDSS data alone. Broadband colours are often regarded as a
straightforward way to distinguish between different kinds of
galaxies. These are available for the SDSS sample, 
but this is a poor discriminant since different Hubble Types can have very similar 
colours (Fukugita et al.~1995), particularly different kinds of spiral galaxies. 
Another discriminant is the light concentration parameter, but this alone cannot be 
used to distinguish different types of late-type galaxies: concentration parameters 
do not depend on scale length if the profiles are exponential. H$\alpha$ emission-line 
strength is yet another discriminant, but there is considerable scatter in the H$\alpha$ 
equivalent width of galaxies of a single Hubble type (\citeauthor{1983AJ.....88.1094K}~1983).

We therefore use the following procedure:
brightward of $M_R=-17.5$, concentration parameters, K-corrected\footnote{The 
K-correction for filter $X$ for a galaxy at redshift $z$ is
defined by the following equation: $K_{X} (z) = 2.5 \log_{10} \left[(1+z) 
                  { {\int_0^{\infty} L_{\lambda} (\lambda^{\prime}) 
                  T_{X} (\lambda^{\prime}) {\rm d}{\lambda^{\prime}}
                  }\over{\int_0^{\infty} L_{\lambda} 
                  ({{\lambda^{\prime}}\over{1+z}}) T_{X} (\lambda^{\prime}) 
                  {\rm d}{\lambda^{\prime}}}}\right].$
It corrects for two effects: (i) the redshifted spectrum is stretched through
the bandwidth of the filter, and (ii) the rest-frame galaxy light that we see 
through the filter comes from a bluer part of the 
spectral energy distribution because of the redshift
\citep{1994AJ....108.1476F}.}
broadband colours, and H$\alpha$ 
equivalent widths were used in conjunction with each other
to classify the SDSS galaxies as early-type, intermediate-type, or late-type using 
local galaxies as templates. Luminosity functions were computed for each. 
The early-type luminosity function was then 
further split into an elliptical luminosity function and an 
S0 luminosity function in such a way that the relative numbers of these 
kinds of galaxies corresponded to that in each magnitude range in the 
{\it Nearby Galaxies Catalogue} (\citeauthor{1988ngc..book.....T}~1988), 
which lists a sample of luminous galaxies within 40 Mpc.
The intermediate luminosity function was split into a Sa luminosity 
function and a Sb luminosity function similarly. 
The late-type luminosity function was split into a Sc, a Sd, and an irregular luminosity 
function. 
Brightward of $M_R=-17.5$, dwarf elliptical galaxies do not seem to exist outside rich clusters
(see e.g.~\citeauthor{1988ARA&A..26..509B}~1988). 
Faintward of $M_R=-17.5$, the luminosity function was split according to the relative numbers of 
the different 
kinds of galaxies in the groups surveyed by Trentham \& Tully (2002) and the Local Group.

This method of computing the luminosity function is motivated by our need to obtain 
stellar mass-to-light ratios and gas masses or galaxies of particular magnitudes and types. 
It forces the SDSS galaxy sample to have properties similar to the local galaxy sample, 
but it generates a luminosity function that is
less susceptible to cosmic variance problems than a luminosity function derived from the 
local galaxy sample alone. Our method is subject to systematic problems if the 
two galaxy samples are very different.
Other authors have used concentration parameters (see e.g.~\citeauthor{2003MNRAS.341...33K}~2003)
or star formation histories derived from the SDSS spectra (Panter et al.~2004) 
to determine the mass-to-light ratios. 
Comparing our results to other values in the literature will be an
important test of our method (see sections \ref{sec:stars},
\ref{sec:gas} and \ref{sec:baryons}).

\begin{figure} 
\begin{center}
\epsfxsize=9truecm\epsfysize=9truecm\epsfbox{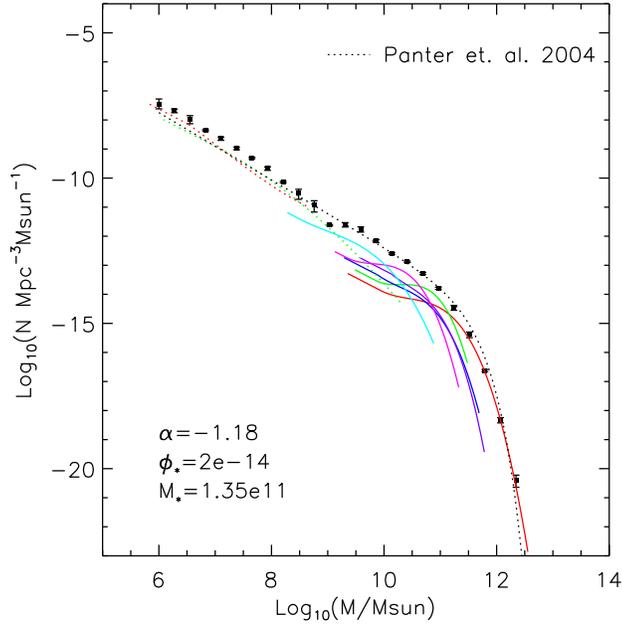}
\end{center}    
  \caption[]
  {The field galaxy stellar mass function split by galaxy Hubble type. The
  data points are for all galaxies, while the lines show spine fits by
  Hubble Type. The lines have the same meaning as in
  Figure 1.
  Bright (massive galaxies) lie to the {\it right} of this plot, while
  faint galaxies lie to the left. Overlaid are parameters for a
    Schechter fit to the total mass function.}
\label{fig:stellarmf}
\end{figure}

Our derived luminosity function agrees very well, both by Hubble type
and integrated, with one obtained using only local galaxies
\citep{1988ARA&A..26..509B}. However, the agreement with a recent
luminosity function obtained from the SDSS alone by
\citet{Blanton:2004zy} is not so good at the faint end (see dotted
black line, Figure \ref{fig:lumfunction}). Their faint end slope,
corrected for incompleteness, is significantly steeper than
ours (they find $\alpha \sim -1.5$; we find $\alpha \sim -1.3$). The
difference could be caused by cosmic variance effects: our particular
patch of the universe may be under-dense. In addition, Schechter fits
to the bright end of the luminosity function (where the Poisson errors
are small) tend to over-predict the number of faint galaxies.

Finally, notice that there is a dip in the luminosity function at $M_R
\sim -17$. A similar dip in the luminosity
function was recently found by \citet{2003Ap&SS.285..191F} suggesting
that it might not simply be the result of incompleteness or poor overlap in
the surveys used (see also \citet{2005MNRAS.tmp...52T} for further
discussion of this feature).

\section{The stellar mass function of field galaxies}\label{sec:stars}

In order to convert the luminosity function to a mass function, we
require the mass-to-light ratio of the stellar populations. 
This will in general be
a function of galaxy type, age, metallicity and even luminosity. 
We determine the mean mass-to-light ratio of ellipticals and assume that this 
may also be applied to the bulge component of S0 and spiral galaxies. We then
determine the mean mass-to-light ratio of discs, irregular galaxies
and dwarf elliptical galaxies. These mean mass-to-light ratios
are then combined using bulge-to-disc ratios, $\mu_B$, 
using the measurements of \citet{1985ApJS...59..115K}. The results are
presented, along with the gas properties of galaxies that we derive in
the next section, in Table 1. In fact, we could do better than just
taking a mean value for each of the Hubble types; in future work we
will use a {\it distribution} of mass-to-light ratios as a function of
luminosity for each Hubble type. This is beyond the scope of this
present work.

\begin{table}
{\vskip 0.75mm}
{$$\vbox{
\halign {\hfil #\hfil && \quad \hfil #\hfil \cr
\noalign{\hrule \medskip}
&   & $\mu_B$ & $M/L_B$  & $m_{\rm HI}/L_B$ & $m_{\rm H_2}/m_{\rm HI}$ &\cr
\cr
\hline
\cr
& E & 1 & 6.8--7.5 & $\sim 0.01$ & $\sim$1 &\cr
& S0+S0/a & 0.64 & 4.8--5.4 & 0.11--0.33 & 1.3--6.7 &\cr
& Sa+Sab & 0.33--0.40 & 3.1--3.6 & 0.078--0.32 & 1.5--5.8 &\cr 
& Sb+Sbc & 0.16--0.25 & 2.1--2.6 & 0.24--0.46 & 2.0--4.2 &\cr
& Sc+Scd & 0.06--0.09 & 1.6--2.1 & 0.46--0.65 & 1.1--1.8 &\cr 
& Sd+Sdm+Sm & $\sim$0 & 1.6--2.1 & 0.62--0.70 & 0.70--1.3 &\cr
& Irr+dIrr & $\sim$0 & 0.9--1.2 & 0.55--0.99 & 0.30--0.58 &\cr
& dE& - & 1.7--4.9 & $\sim$0 & - &\cr
\cr
\hline
\cr
\noalign{\smallskip \hrule}
\noalign{\smallskip}\cr}}$$}
\caption[]{Galaxy properties as a function of Hubble Type. }
\label{tab:galprops}
\end{table}

\subsection{A note on random v.s. systematic errors}\label{sec:errors}
\noindent
Stellar masses depend on both the galaxy luminosity function and the
mass-to-light ratios of the stellar populations in the different kinds of
galaxies. The errors in the former are random and are determined by
counting statistics. The errors in the latter are systematic and come
from our lack of knowledge about input parameters like metallicity and
stellar IMF in the stellar population models used to calculate the
mass-to-light ratios. We can get around this problem to some extent by
comparing these mass-to-light ratios with those derived from dynamical
measurements; we may then quote a range of
mass-to-light ratios determined from these two methods. In this paper
we take this pragmatic approach. However, the reader should be aware
that dynamical measurements have their own systematic uncertainties
and so it is not clear that we can really quantify all of our
systematics in this way. As a conservative estimate, we suggest that
the total systematic error may be as large as $\sim 50$\% (compared with our
quoted errors of $\sim 10$\%). In future, we may hope to better
quantify these systematics through spectroscopy of individual stars,
deep photometry and improved modelling.

\subsection{Ellipticals and bulges}
\noindent
Two methods were employed here. The first uses
stellar population synthesis models to convert an age, metallicity and colour into
a $B$-band mass-to-light ratio $M/L_B$. We use the models of
\citet{2003MNRAS.344.1000B} with an initial {\it stellar} mass
function (IMF), $\eta(M)$\footnote{The stellar IMF is defined
  such that $\eta(M)$ is the number of stars in the interval $M$ to
  $M+dM$. 
  ${\rm M}_\odot=1.989 \times 10^{30}$ kg is the mass of the sun.},
 taken from \citet{2001MNRAS.322..231K}: 

\[
\eta(m)  =\left\{ \begin{array}{l@{\quad : \quad}l}
 25 \, \eta (1) \,m^{-0.3}  & 0.01 \leq m < 0.08 \\
 2  \, \eta (1) \,m^{-1.3}  & 0.08 \leq m < 0.5 \\
    \, \eta (1) \,m^{-2.3}  & 0.5 \leq m < 100
\end{array}\right.
\]
\begin{equation}
\label{eqn:kroupaimf}
\end{equation}

\noindent where $m$ is in units of solar masses ${\rm M}_\odot$.
The results are
very sensitive to the IMF, since
arbitrarily many low-mass stars can be included with no change to the
measured colour, metallicity or age of a galaxy.

This IMF has been {\it empirically} determined from
deep observations of local field stars
and of young star clusters. It is denoted
the `universal IMF' since it appears to be the same across the
enormous range of scale, environment and epoch in which it has been
determined \citep{2001MNRAS.322..231K}. 
This IMF is also attractive in that it is the end result of star
formation that exhibits a Salpeter IMF whilst in 
progress (\citeauthor{2003ApJ...598.1076K}~2003).

Using the stellar population models of \citet{2003MNRAS.344.1000B},
adopting the IMF above, and assuming a mean age and
metallicity for ellipticals of $t_{E} \simeq 12 \pm 2$ Gyr and $Z =
2 \, {\rm Z}_\odot$\footnote{The
  metallicity, $Z$, is the ratio of the
  total mass in heavy elements, $M_h$, to the total baryonic mass
  \citep{1998gaas.book.....B}.} respectively, we obtain $M/L_B =
9.81-11.05$\footnote{This is in units of $M_\odot L_{\odot,B}^{-1}$ -
  solar masses per B-band solar luminosity, which we use throughout
  this paper.}. Note that fitting a power law
to the stellar population models of \citet{2003MNRAS.344.1000B} gives
$M/L_B \sim 6.7 ({\rm Z}/{\rm Z}_\odot)^{0.58}$. Thus the results are
sensitive both to the choice of IMF, and to the choice of mean
metallicity for the ellipticals.

We may also obtain the mass-to-light ratio of ellipticals from
dynamics. While there is significant evidence for dark matter in
ellipticals, interior to the majority of the stellar light the dark
matter component is small (\citeauthor{1991MNRAS.253..710V}~1991 and
\citeauthor{2000AAS..144...53K}~2000) so the stellar velocities (as measured
from their relative Doppler shifts) may be used to calculate the total
gravity produced by masses of the stars. 

A recent study of 21 elliptical galaxies by
\citeauthor{2000AAS..144...53K}~2000 gives $M/L_B = 6.78-7.54$ which
is a factor $\sim 0.7$ of the stellar population value. This
is surprising since the presence of dark matter should tend to
make the stellar populations value {\it smaller} than the dynamical
estimate rather than the other way round.
We favour the dynamical estimate because modelling stellar
populations is extremely difficult and errors of the order 2 are not
uncommon \citep{1996ApJ...457..625C}. Perhaps more importantly, the
`universal' IMF of Kroupa (2001) has not been
explicitly verified for elliptical galaxies and so may not be the
correct one to use
\citep{1997MNRAS.288..161J}. \citet{2003PASP..115..763C} suggest an
IMF which gives a much better agreement between stellar population
models and dynamical masses for elliptical galaxies, however similar
agreement may be obtained by using the Kroupa (2001) IMF and assuming
a mean metallicity for ellipticals of solar, rather than twice
solar. The double degeneracy of metallicity and IMF for the
ellipticals leads us to favour the dynamical value for the mass to
light ratio. This has its own systematic limitations: the internal
velocity dispersions of ellipticals are unknown since only the line of
sight velocities may be measured; and the presence of dark matter may
complicate things at large radii. Future dynamical mass measurements,
improved stellar population modelling, deep photometry of nearby
galaxies and good spectra of individual stars will all help to pin
down these systematic uncertainties in the future.

\subsection{The mass-to-light ratio of discs}
\noindent
Following \citet{1998ApJ...503..518F} we compile a mean mass-to-light
ratio for discs from three independent methods. The first is from
measurements of the column density of stars in the solar
neighbourhood, $\Sigma_*=27-40$M$_\odot$pc$^{-2}$
(\citeauthor{1996ApJ...465..759G}~1996 and
\citeauthor{1989MNRAS.239..605K}~1989). Combining this with the local luminosity 
surface density, $\Sigma_{L_B} \simeq 18$M$_\odot$pc$^{-2}$
\citep{1980ApJS...44...73B} gives $M/L_B=1.5-2.2$.
The second method uses the stellar population synthesis models of \citet{1981ApJ...250..758T} and
\citet{2004MNRAS.347..691P}. These are on a firmer footing than for
the elliptical galaxies since the IMF has been explicitly measured for
spiral galaxies \citep{2001MNRAS.322..231K}. These give
$M/L_B = 1.1-1.9$ using the IMF given in equation \ref{eqn:kroupaimf}. 
The third method uses a dynamical mass estimate for the stars 
(\citeauthor{1999MNRAS.309..923S}~1999). This gives $M/L_B =1.11$.

The mean of all of these values gives 
$M/L_B = 1.24-1.7$. We adopt this value here.
This value was then corrected for internal extinction in the galaxies,
using the corrections of \citep{1985ApJS...58...67T} and a field
sample inclination distribution equivalent to that seen in the Ursa
Major Cluster \citep{1996AJ....112.2471T}. A similar value for $M/L_B$
was recently obtained dynamically from a compilation of spiral galaxies by
\citet{2004ApJ...609..652M}. This further strengthens our confidence
in this value.
 
\subsection{The mass-to-light ratio of dwarfs and irregulars} 
\noindent
The mass-to-light ratios for the irregulars 
(and dwarf irregulars) may be adapted
from the above value for discs corrected for their younger age and
consequent bluer colour  \citep{1998ApJ...503..518F}:
$M/L_B=0.9-1.24$.
 
Recent observations of the Ursa Minor Local Group dwarf elliptical
(dE)\footnote{Ursa Minor or UMi is usually labelled as a dwarf
  spheroidal galaxy (dSph) which is an alternate name for a dwarf
  elliptical galaxy.}
galaxy by \citet{2002NewA....7..395W} suggest that dE stellar
populations are very similar to the Milky Way globular clusters. As
such, we use the dynamical mass-to-light ratio of 
Galactic globular clusters (\citeauthor{1993ASPC...50..357P}~1993) for
dE galaxies: $M/L_B=1.7-4.9$.

We find excellent agreement between our derived stellar mass function
and that of \citet{2004MNRAS.355..764P} (see the black dotted line in Figure
\ref{fig:stellarmf}). They also use the SDSS data, but derived stellar masses
from star formation histories constrained by the spectra of
the galaxies. This gives some weight to the validity of our method of
classifying galaxies by Hubble type and then determining masses.
Our data gives $\Omega_* = 0.0028 \pm 0.0003$, whereas Panter et al.~find
$\Omega_* = 0.0034 \pm 0.0001$.
 
\section{The gas mass function of field galaxies}\label{sec:gas}

\begin{figure} 
\begin{center}
\epsfxsize=12truecm\epsfysize=6truecm\epsfbox{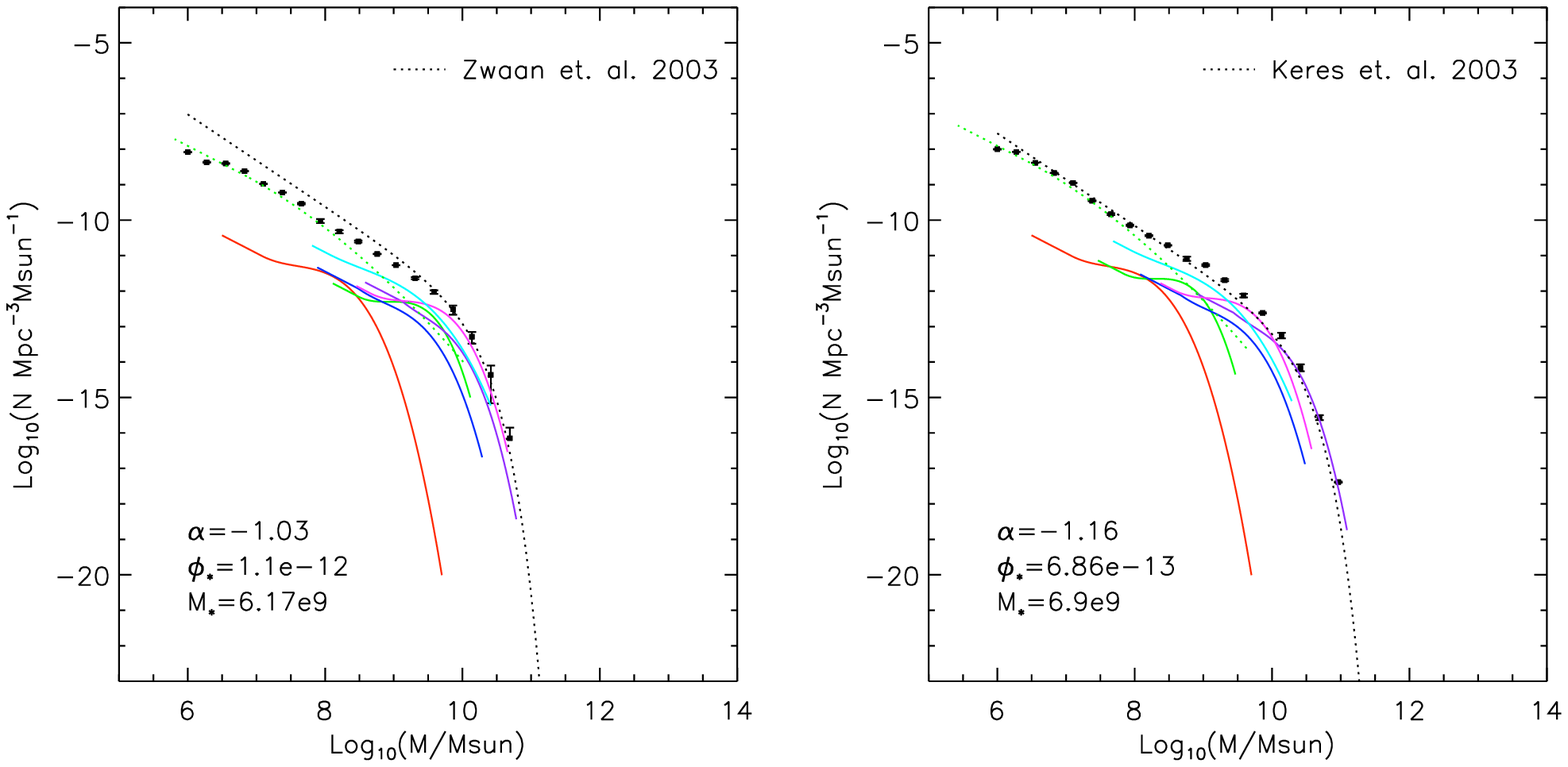}
\end{center}
  \caption[]
  {The field galaxy HI (left) and H$_2$ (right) gas mass functions split
  by galaxy Hubble Type. The
  data points are for all galaxies, while the lines show spine fits by
  Hubble type. The lines are as in Figure 1. Overlaid are parameters for a
    Schechter fit to the total mass function.}
\label{fig:higasmf}
\end{figure}

The gas mass function of local galaxies is presented in Figure
\ref{fig:higasmf}.

\subsection{Atomic hydrogen (HI)}
\noindent
A large sample of HI (atomic) gas measurements of nearby galaxies was compiled  
recently as part of the Australian HIPASS survey, using the Parkes radio telescope. 
Measurements of the 1000 brightest galaxies in this sample were recently published 
(Koribalski et al.~2004). 

In order to measure the total amount of gas in galaxies in the Universe, we need to determine how much gas the 
galaxies contain per unit optical luminosity. We can then get 
the gas mass function from the luminosity function. 
We computed the atomic gas mass per unit luminosity for each Hubble Type as follows.
We compiled a sample of galaxies with velocities between 2000 and 3000 km/s 
that were observed at optical wavelengths and also in HI in the HIPASS survey. For each Hubble Type,  
we used this sample to determine the ratio between the atomic gas mass and the optical luminosity. These numbers are given in Table 2. The wide ranges reflect 
the conservative error estimates that we are forced to use because of the intrinsic scatter in the ratio for galaxies of a given
Hubble Type and because the errors are systematic (see also section 3\ref{sec:errors}).

The HI mass function that we derived is very similar to the one derived by HIPASS (Zwaan et al.~2003) at the high-mass end, 
but falls significantly below theirs at the low-mass end. 
This difference can be attributed to the difference in sample selection and scaling --
the HIPASS sample is selected by HI mass but ours is selected and scaled by optical luminosity. They therefore include many 
systems that we do not, as well as including gas in the outer parts of galaxies which is not included in our analysis -- 
a galaxy that we think has mass $M$ of atomic gas may really have $2M$, so our points should be shifted towards the right if we 
wish to include all the atomic gas in the Universe. Both of these effects could be more important at low masses -- the highest mass 
HI galaxies will invariably be in optical surveys, and all scaling relations were derived using the properties of these galaxies..
This difference also accounts in our different values of $\Omega_{\rm
  HI}$: HIPASS finds $\Omega_{\rm HI}=(4.1 \pm 0.6) \times 10^{-4}$,
whereas we get $\Omega_{\rm HI}=(2.9 \pm 0.6) \times 10^{-4}$.

\subsection{Molecular hydrogen (H$_2$)}
\noindent
Similarly, we determine the relationship between galaxy Hubble Type and molecular gas 
mass using the data of \citet{1995ApJS...98..219Y}. The results are also presented in Table 2. 
This survey measures CO line luminosities, which are converted to gas masses
using the empirical formula of \citet{1991ARA&A..29..581Y}: 
\vskip 1pt \noindent $M_{\rm H_2} = 12000 \, S_{\rm CO} D^2 {\rm M}_\odot$, 
where $S_{\rm CO}$ is the CO flux in Jy km/s and $D$ is the
distance to the galaxy in Mpc.
 
We find a very similar molecular gas mass function to that of Keres et al.~(2003), 
who presented this function for a far-infrared 
(FIR)-selected sample, also derived from the Young et al.~(1995) sample. This is consistent with 
the idea that most of the currently observed molecular gas in the
Universe is in normal galaxies, which are the same objects as the FIR
sources.

\section{The baryonic mass function of field galaxies}\label{sec:baryons}

\begin{figure} 
\begin{center}
\epsfxsize=9truecm\epsfysize=9truecm\epsfbox{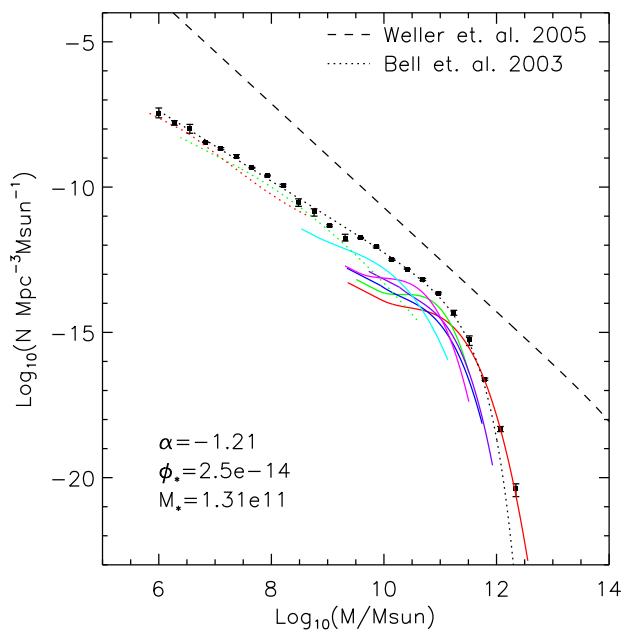}
\end{center}
  \caption[]
  {The field galaxy baryonic mass function. The
  data points are for all galaxies, while the lines show spine fits by
  Hubble Type. The lines are as in Figure \ref{fig:stellarmf}.
The CDM mass spectrum from the numerical simulations of  
\citet{Weller:2004sc} is also shown. Overlaid are parameters for a
    Schechter fit to the total mass function.}
\label{fig:baryonicmf}
\end{figure}

The baryonic mass function is presented in Figure \ref{fig:baryonicmf}. 
This is the sum of the functions in the three previous Figures, 
where the masses in the gas functions are multiplied by 1.33 to take into
account the presence of helium. The error bars are small enough that
Schechter or any other analytical fits to the data are formally poor.
However, we show our best fit values on the plot.

In Table 3, we present the total mass density 
of the Universe in baryons within galaxies, in different forms.
Typical errors are 10\% (but see section 3\ref{sec:errors}). For the stellar 
masses the uncertainty mainly comes from our lack of knowledge of the mass-to-light ratios. For the 
gas components the error primarily comes from the uncertainty in the scaling of optical luminosity to 
gas mass for galaxies of a given Hubble Type.

About 8\% of the baryons in galaxies are in atomic gas and about 7\%
are in molecular gas. In the Milky Way about 8\% of the baryons are in
atomic gas and a further 2\% are in molecular gas.

A comparison with the CDM mass function is also shown in Figure 4. 
The normalisation for the baryonic component in galaxies is very much 
lower, reflecting the fact that galaxy masses are dominated by dark matter, not baryons. 
Additionally, the baryonic mass function has a very different shape from the dark mass function. 
This suggests that the collapse of baryons into galaxies and the ability
of the galaxies to retain the gas once
star formation has begun is very much a scale-dependent process. 
At the very low mass end, where galaxies are heavily dark-matter
dominated, the baryonic mass function is still very much shallower
than the CDM mass function -- this is the classical missing satellites
problem \citep{1999ApJ...524L..19M}. At the high mass end it is also
deficient. This could be caused by feedback from centrally active
black holes within galaxies \citep{1998AA...331L...1S}.

Finally, notice that there is a slight bump in the baryonic mass
function at $\sim 10^9$M$_\odot$. This corresponds to a similar dip in
the stellar mass function (see Figure \ref{fig:stellarmf}) and a dip in the
luminosity function at $M_R \sim -17$ (see Figure
\ref{fig:lumfunction} and section \ref{sec:luminosity}). The dip is much
less significant in the 
baryonic mass function than in the stellar mass function and the
luminosity function. It is possible in all cases that this dip is a
statistical artifact: its significance is not high. However it is
interesting that it occurs at the transition point in galaxy type from
ellipticals and spirals to dwarfs. The fact that it is less pronounced
in the full baryonic mass function than in the stellar mass function
may then simply highlight the fact that the gas mass fraction in dwarf
irregulars is much larger than in ellipticals and spirals.

\begin{table}
\begin{center}
\setlength{\arrayrulewidth}{0.5mm}
\begin{tabular}{lllll}
\hline
& $\Omega_{b}$ & $\Omega_{*}$ & $\Omega_{HI}$ & $\Omega_{H_2}$ \\
\hline
E         & 0.00064 & 0.00064 & $\sim 0$ & $\sim 0$ \\
S0        & 0.00073 & 0.00068 & 0.00003 & 0.00001 \\
Sa+Sab    & 0.00036 & 0.00032 & 0.00001 & 0.00002\\
Sb+Sbc    & 0.00056 & 0.00040 & 0.00004 & 0.00008\\
Sc+Scd    & 0.00072 & 0.00047 & 0.00010 & 0.00008\\
Sd+Sdm+Sm & 0.00037 & 0.00021 & 0.00007 & 0.00005\\
Irr+dIrr  & 0.00013 & 0.00007 & 0.00003 & 0.00001\\
dE  & 0.00002 & 0.00002 & $\sim 0$ & $\sim 0$ \\
\hline
total         & 0.0035 & 0.0028 & 0.00029 & 0.00026 \\
\end{tabular}
\end{center}
\caption[]{The baryonic mass density of galaxies. Typical errors are
  $\sim 10$\%. A discussion of the larger systematic errors is given
  in section 3\ref{sec:errors}.}
\label{tab:omegawefind}
\end{table}

There are a number of baryonic mass components which we have
not included in the above analysis because they comprise only a tiny
fraction of the total mass in baryons within galaxies.\\
\\
\noindent
{\bf Field halo stars:} In the Milky
Way and M31 there is a faint spheroidal stellar population of stars
which lie neither in the disc nor bulge components of the galaxy
\citep{1989ARAA..27..555G}.
While the total mass of this component is
quite uncertain and estimates range from $\sim 10^8$
\citep{1998gaas.book.....B} to $\sim 10^9$
M$_{\odot}$ 
\citep{2002ARA&A..40..487F}, this is still only $\sim 0.1-1$\% 
of the total baryonic mass and, therefore, a negligible Galactic mass
component. Because of the low surface density of these
halo stars, they are difficult to resolve in galaxies much further
away than M31. Perhaps other galaxies have a much more significant
stellar halo and we are missing an important baryonic mass
component. 
This is probably not the case, however, because stars from these
halos would generate a high extragalactic background light in galaxy
clusters, which is not observed \citep{Zibetti:2005tt}\footnote{In
  fact an argument of this sort requires some
  care. \citet{Zibetti:2005tt} find that $\sim 30$\% of the stars in
  clusters can be attributed to background light of some
  sort. However, $\sim 20$\% of this is associated with the large
  central cluster galaxy. It is likely that these these stars are
  tidal debris left over from infalling galaxies; they would have
  formed within discs or bulges, rather than stellar halos. We take
  what we feel is a conservative upper bound from these cluster
  measurements of 10\% for the contribution from stars
  which formed within stellar halos.}.\\
\\
\noindent
{\bf MACHOs and stellar remnants:}
Limits can be placed on massive compact halo objects (MACHOs) of any form, 
which may or may not be made of baryonic matter, from microlensing experiments (e.g.~Alcock et al.~2000). 
Current data (Afonso et al.~2003) suggests that the contribution of MACHOs to the 
total baryonic mass of the Galaxy may be small, and we neglect this contribution here. 
A more stringent constraint can be placed on the total mass in MACHOs that are the endpoints of stellar evolution 
(neutron stars and black holes) due to the lack of large amounts of heavy elements in the Universe   
(e.g.~\citeauthor{2001idm..conf..213F}~2001). Fukugita \& Peebles
(2004) estimated the cosmological density in the remnants to be
$\Omega_{\rm remnants} \sim 0.00001$, which is far less than the sums
in Table 3, and we ignore this contribution here.\\
\\
\noindent
{\bf Hot gas:}
There are two main hot ionised gas components in galaxies: a 
component associated with HII regions, which is heated by ionising
radiation from O and B stars, and an extended component which
is associated with the dark matter halo. The mass of the component associated
with HII regions is negligible ($\sim$ a tenth of 
the molecular gas mass) for both early and late type galaxies
(e.g.~\citeauthor{2001MNRAS.328..461O}~2001,
\citeauthor{2003ARAA..41..191M}~2003 and
\citeauthor{1999ASPC..163...55G}~1999). The extended 
component is what we refer to later as the warm/hot intergalactic
medium or WHIM. Some of this gas is likely bound to galaxies; much of
it presumably also resides in the intergalactic medium. This
component, which we do not count as part of galactic baryonic matter,
is probably the dominant form of baryons in the Universe.\\
\\
\noindent
{\bf Dust:}
In the Milky Way, a dust-to-gas ratio by mass is about 1/200
\citep{1989ARAA..27..555G}. 
There does not seem to be any evidence for this ratio being different in the majority of external normal galaxies 
(see e.g.~the review by Young and Scoville 1991), and we neglect the
contribution of dust to the baryonic mass density of the Universe.

\section{The contribution to $\Omega_b$}\label{sec:omega} 

The value that we get of $\Omega_{\rm b, gal} \sim 0.0035$ in galaxies is only about one-tenth of the 
value of the baryonic density of the universe $\Omega_{\rm b} \sim 0.05$  derived from measurements
of light element abundances (\citeauthor{2004ApJ...600..544C}~2004)
and from the shape of the angular power spectrum of the cosmic microwave background \citep{2003ApJS..148..175S}. 
This opens up the important question: if they are not in galaxies, where do 90\% of the baryons in the Universe reside?

Hydrodynamic simulations (\citeauthor{1999ApJ...514....1C}~1991) suggest that most of these baryons are in 
a warm/hot intergalactic medium (WHIM), that is highly ionised. This medium is very difficult to detect, but there might be indirect 
evidence for its existence, from observations of OVII X-ray absorption along the lines of sight to high redshift quasars
(e.g.~\citeauthor{2003ApJ...586L..49F}~2003). Directly detecting the WHIM and will be a major area of observational study over the next few years.

\section{Conclusions}\label{sec:conclusions}

The main conclusions from this work are as follows:

\vskip 5pt \noindent 
1.~The baryonic density of the universe that resides in galaxies is
$\Omega_{\rm b,gal} = 0.0035 \pm 0.0003$.
This is far less than the value of $\Omega_{\rm b} = 0.05$ inferred from the CMB or from BBN. 
Most of the baryons in the Universe do not reside in galaxies and
probably reside in the warm/hot intergalactic medium (WHIM). Direct detection
of the WHIM will form one of the major challenges for astronomy in the
next decade.

\vskip 5pt \noindent
2.~Most of the baryons in galaxies are in stars, not gas, and about one-half of these stars are in early-type galaxies.
We derive the value of $\Omega_*=0.0028 \pm 0.0003$, assuming stellar mass-to-light ratios derived from 
population synthesis models, a Kroupa IMF in discs and irregular galaxies and 
stellar mass-to-light ratios derived from dynamical measurements in bulges and elliptical galaxies.

\vskip 5pt \noindent
3.~About 15--20\% of the baryons in galaxies are in gas. Of this, about one half is in molecular gas.

\vskip 5pt \noindent
4.~About 30\% of the atomic gas seen by HIPASS is not present in our atomic gas mass function. 
We attribute this to gas in galaxies that are missing in the sample that we used to define 
a scaling between optical luminosities and gas masses.\\
\\ 
\noindent
The main uncertainties in this work arise from the systematic errors
in determinations of the stellar and gas mass-to-light ratios. The
errors we quote throughout this paper are small, since they derive
from the range in mass-to-light ratios determined from different
methods. Such errors do not take account of systematic errors {\it within}
each mass determination which are much harder to quantify and may be
as large as 50\%. Spectroscopy, deep photometry and improved modelling
will all help to pin down and reduce such systematics in the
future. An accurate baryonic mass function is already beginning to
form one of the observational pillars of modern cosmology.

\section{Acknowledgements}
We would like to thank the three anonymous referees, Stacy McGaugh
and Gilles Chabrier for their useful comments.

\bibliographystyle{mn2e}
\setlength{\baselineskip}{0mm}
\setlength{\parskip}{0mm}
\setlength{\parsep}{0mm}
\setlength{\parindent}{0mm}
\setlength{\listparindent}{0mm}
\setlength{\itemsep}{0mm}
\bibliography{refs}

\end{document}